\documentclass[aip,
 rsi,
 amsmath,amssymb,
 reprint,%
]{revtex4-1}

\usepackage{graphicx}% Include figure files
\usepackage{dcolumn}% Align table columns on decimal point
\usepackage{bm}% bold math
\usepackage[utf8]{inputenc}
\usepackage[T1]{fontenc}
\usepackage{mathptmx}
\usepackage{natbib}
\usepackage{siunitx}

\begin{document}

\preprint{AIP/123-QED}

\title[]{Composite Pressure Cell for Pulsed Magnets}
% Force line breaks with \\

\author{Dan Sun}
\email{ustsundan@gmail.com}
\affiliation{National High Magnetic Field Laboratory, Los Alamos National Laboratory, Los Alamos, NM, 87544, USA}%Lines break automatically or can be forced with \\
\author{Martin F. Naud}
\affiliation{National High Magnetic Field Laboratory, Los Alamos National Laboratory, Los Alamos, NM, 87544, USA}
\affiliation{Mechanical Engineering, University of Waterloo, Waterloo, ON, N2L3G1, Canada}
\author{Doan N Nguyen}
\affiliation{National High Magnetic Field Laboratory, Los Alamos National Laboratory, Los Alamos, NM, 87544, USA}
\author{Jonathan Betts}
\affiliation{National High Magnetic Field Laboratory, Los Alamos National Laboratory, Los Alamos, NM, 87544, USA}
\author{John Singleton}
\affiliation{National High Magnetic Field Laboratory, Los Alamos National Laboratory, Los Alamos, NM, 87544, USA}
\author{Fedor Balakirev}%
\affiliation{National High Magnetic Field Laboratory, Los Alamos National Laboratory, Los Alamos, NM, 87544, USA}%

\date{\today}% It is always \today, today,

\begin{abstract}
Extreme pressures and high magnetic fields can affect materials in profound and fascinating ways.  However, large pressures and fields are often mutually incompatible; the rapidly changing fields provided by pulsed magnets induce eddy currents in the metallic components used in conventional pressure cells, causing serious heating, forces, and vibration. Here we report a diamond-anvil-cell made mainly out of insulating composites that minimizes inductive heating while retaining sufficient strength to apply pressures of up to 8~GPa. Any residual metallic components are made of low-conductivity metals and patterned to reduce eddy currents. The simple design enables rapid sample or pressure changes, desired by pulsed-magnetic-field-facility users. The pressure cell has been used in pulsed magnetic fields of up to 65~T with no noticeable heating at cryogenic temperatures. Several measurement techniques are possible inside the cell at temperatures as low as 500~mK. 
\end{abstract}

\maketitle

\clearpage

\section{Background}

Studies of electronic and magnetic materials under extreme conditions such as high pressure~$(P)$, high magnetic field~$(B)$, and low temperature, have provided great insights into the quantum-mechanical mechanisms that determine their key properties \cite{national2013high,drozdov2019superconductivity,mozaffari2019superconducting,clune2020developing,helm2020non,goddard2002superconductivity,bangura2007angle,ghannadzadeh2013evolution}. The combination of pressure and magnetic fields has frequently been used to observe phenomena that are unknown under ambient conditions \cite{national2013high,drozdov2019superconductivity,mozaffari2019superconducting,goddard2002superconductivity,bangura2007angle,ghannadzadeh2013evolution}. Pressure compresses or distorts crystalline lattices and thus varies the overlap of the electron orbitals in a systematic and reversible manner \cite{goddard2002superconductivity,bangura2007angle,eremets1996high,singleton2000studies}. If the pressure is truly hydrostatic, it changes physical properties whilst preserving crystal symmetry \cite{clune2020developing,bangura2007angle,ghannadzadeh2013evolution,singleton2000studies}; in contrast to when alloying is used to the same end, little disorder is introduced. Examples of effects observed and characterized using a combination of high pressures and magnetic fields include systematic changes in band structure in low-carrier-density systems~\cite{goddard2002superconductivity,bangura2007angle,singleton2000studies}, the creation of phases such as novel superconducting states~\cite{national2013high,drozdov2019superconductivity,mozaffari2019superconducting,goddard2002superconductivity}, and tuning towards quantum-critical points~\cite{helm2020non,singleton2000studies}. Therefore techniques for applying pressure in the highest possible nondestructive pulsed magnetic fields $(\sim 100~{\rm T})$ are highly desirable~\cite{national2013high}. 

Measurements up to several GPa are usually performed in bulky metallic pressure cells which rely on the high tensile strength of nonmagnetic steel or copper-beryllium alloys \cite{eremets1996high}. However, combining such cells with pulsed magnetic fields presents insurmountable challenges. First, in such magnets, the large values of ${\rm d}B/{\rm d}t\sim 10$~kT/s \cite{nguyen2020comprehensive,michel2020design} will induce substantial eddy currents in any metallic cell components, leading to heating, mechanical forces and vibrations. Second, in order to maximize the field whilst remaining below the tensile strength of available coil materials, the inner diameter of the pulsed magnet coil is kept very small \cite{nguyen2020comprehensive,michel2020design}. Only a few groups around the world have attempted to tackle these issues and have made pressure cells for pulsed magnets \cite{nardone2001hydrostatic,graf2011nonmetallic,wang2014use,miyake2015development,braithwaite2016pressure,mozaffari2019superconducting}.

Nardone \textit{et al}.~\cite{nardone2001hydrostatic} built a pressure cell for a long-pulse magnet with a peak field of 33~T and pulse duration $\sim 1$~s. The cell has commercial yttria-doped zirconia anvils and the body is made of TA6V, a Titanium Alloy. Pressures as high as 1~GPa were reported. The outer diameter of the cell is 18~mm, which exceeds the 15~mm and 10~mm bore sizes of the 65~T and 100~T magnets, respectively, available at Los Alamos \cite{nguyen2016status}. Owing to the use of metallic materials, the long field pulse causes a 0.05~K increase in temperature at 1.8~K. However, ${\rm d}B/{\rm d}t$ for the Los Alamos pulsed magnets is a factor $\sim 10$ greater than that of the magnet in Ref.~\cite{nardone2001hydrostatic}, so that the heating would be $\sim 100 \times$ larger.

Graf~\textit{et al}.~designed a turnbuckle pressure cell for use in pulsed fields~\cite{graf2011nonmetallic}. The cell's reported pressure limit is 11.5 GPa using anvils with 0.54 mm culets~\cite{Tozer2020}. Cell dimensions permit it to be rotated inside a pulsed magnet. The cell is completely non-metallic. The gaskets are made from a heavily loaded diamond-epoxy mix and reinforced with Zylon HM fiber. The cell body is made from Parmax 1200, which has high tensile strength (203~MPa) and compressive strength (351~MPa). Unfortunately, the manufacture of Parmax~1200 has ceased; worse, it is not compatible with solvents and glues that are frequently used in the laboratory. Moreover, the turnbuckle design of the cell causes the anvils to move by twice the displacement of the screw thread, making the fine control of pressure difficult~\cite{graf2011nonmetallic}. 

Wang~\textit{et al}.~\cite{wang2014use} followed the turnbuckle design and employed carbon-fibre-reinforced polyether ether ketone (PEEK - also known as 90HMF40) as the body material. With 0.8 mm culets, the cell can generate a pressure of 5.6~GPa, with the maximum value limited by the plastic deformation of the end nuts. The cell was redesigned using custom-fabricated diamonds as the anvils; however, the modified cell is too large to rotate inside the pulsed-field sample space.

Miyake~\textit{et al}.~\cite{miyake2015development} used polybenzimidazole (PBI) to make a cell for a pulsed magnet at the Institute for Solid State Physics (University of Tokyo). The cell is much larger than the previous examples discussed (outer diameter 20~mm, length 30~mm) and it utilizes stainless-steel gaskets. In a test at low temperature,  hysteresis between measurements on the up- and downsweeps of the pulsed field was unnoticeable. However, the material used for the test has only a small temperature dependence below 20~K, so that the true heating effect is unknown.

Braithwaite~\textit{et al}.~\cite{braithwaite2016pressure} designed a pressure cell for resistivity measurements in a 60~T pulsed magnet which reaches a pressure of 4~GPa. The sample space of the gasket has a diameter of 1.2~mm, big enough to hold two four-wire experiments. However, with an outer diameter of 15~mm, the cell is too big for the sample space available in the Los Alamos pulsed magnets~\cite{nguyen2016status}. The cell body is made of MP35N alloy. The temperature increase at 1.5~K is less than 1~K in a ${\rm d}B/{\rm d}t\sim 5-6$ times smaller than in our magnets; heating could therefore be an issue in the Los Alamos pulsed magnets. A slightly larger metallic DAC cell was employed in pulsed field photoluminescence studies up to 10 GPa, although sample heating by eddy currents was not investigated~\cite{millot2008high,millot2010electronic}. 

A miniature all-metal diamond anvil pressure cell was employed in our 65~T pulsed magnets to investigate a high-$T_\text{c}$ hydride superconductor at 160~GPa~\cite{mozaffari2019superconducting}. The cell is made of CuTi alloy and has an outer diameter of 8.8~mm. The cell heated by 5~K from an initial 200~K {\it after} the field pulse; even larger increases were observed at lower temperatures.  Below 50~K the cell heats up {\it during} the magnet pulse, making the sample temperature as a function of field uncertain. 

\section{Pressure-cell design}

To satisfy the demanding requirements of the pulsed-field environment, the cell must be small, strong, and present a minimal cross-sectional area of metal to the field. To address these challenges, we chose a composite design with non-metallic seats for the diamond anvils and non-metallic gaskets to avoid ${\rm d}B/{\rm d}t$-induced eddy-current heating. However, to avoid issues related to the limited strength of plastic threads, the anvils are compressed together via thin metal rods made from a high-strength titanium alloy; this has the advantage of a relatively high resistivity. The design reaches a pressure $\approx 8$~GPa on a 0.8~mm diameter culet. With an outer diameter of 9.2 mm and a length of 26 mm, the cell fits in the double wall $^3$He fridges commonly used in Los Alamos. 

\subsection{Anvil seats}
\begin{figure}
\includegraphics[width=0.8\linewidth]{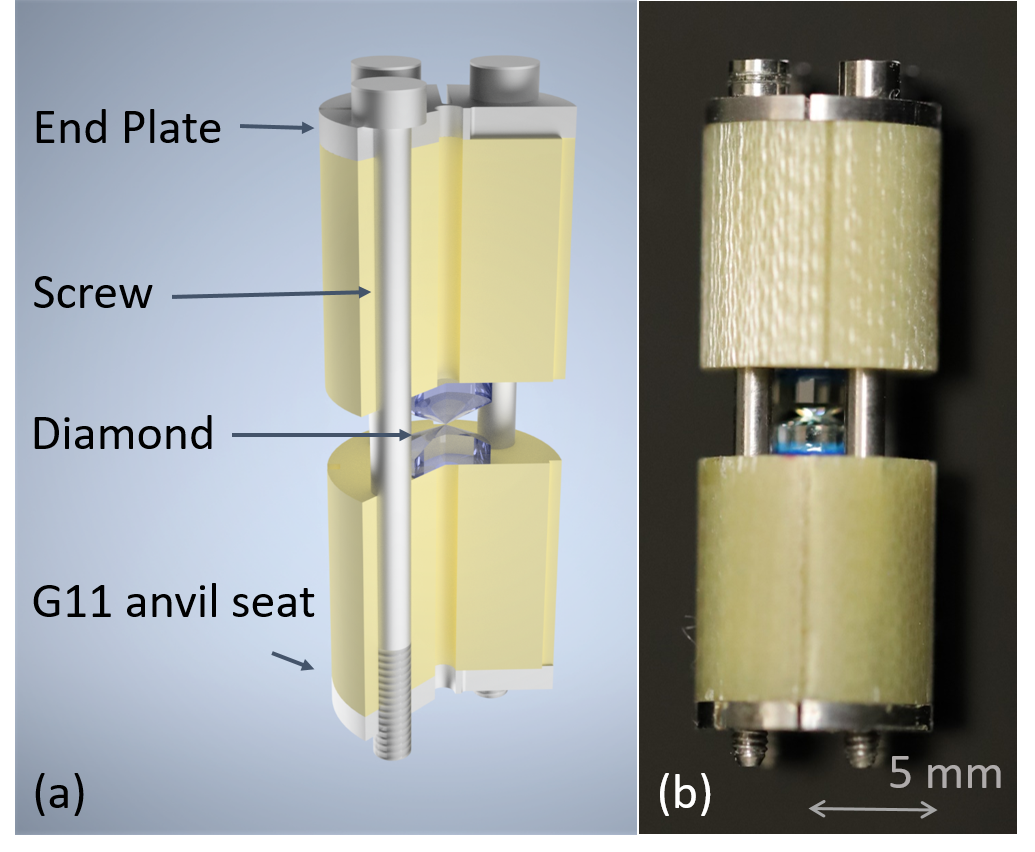}
\caption{\label{fig:cell_drawing} (a)~Drawing of the cell. From top to bottom, the cell consists of an end plate without threads, one G11 anvil seat, two diamonds, another G11 seat and another end plate with threads. All of the above parts are aligned and locked using four screws. For clarity, the gasket between the diamonds is not shown. (b)~A~photo of the pressure cell. }
\end{figure}

The anvil seats are designed to ensure a quick turn-around rate, desirable for pulsed-field-facility users with limited time on-site. The cell uses G11 fiberglass as the anvil-seat material, as shown in Fig.~\ref{fig:cell_drawing}. This reinforced fiberglass is one of the strongest readily available plastic materials, and is easy to machine.  

There are four guide holes for the screws in the seats. The screws are made to slip-fit in elongated holes, with only a short threaded part at the end to serve as both the alignment and the tightening mechanism.  To ensure good alignment of the anvils, the seats' working surfaces are precisely machined perpendicular to the alignment screws. The diamonds are fixed to the cell body with Stycast 2850, for easy assembly of the cell. One diamond is fixed to the seat first, with a fixing jig. The other diamond is then placed between the first diamond and the other seat, and then the cell is assembled. The translational overlap of the culet is examined by a microscope, to a precision better than 20 $\mu$m, with the uncertainty coming from the 50 $\mu$m bevel.  The parallelity of the two culet surfaces is ensured by the precise machining of the seat surfaces and the surfaces of the diamonds. 
The good alignment of the diamonds is tested by the appearance of Newton's fringes~\cite{brooker2003modern} through the center hole in the seats before and after the Stycast is cured. 
Fewer than 4 fringes are allowed in the alignment, hence the mis-alignment angle is less than $\theta \approx 4\times (500~{\rm nm}/2)/800~\mu {\rm m} \approx 10^{-3}$ rad, where we use 
500~nm as the central wavelength of visible light and $800~\mu$m as the diameter of the culet. To route the signal wires safely through the restricted cryostat space, four vertical grooves are provided on the perimeter of the seats. To ensure a large sample space for a variety of measurements, we use diamonds with 0.8~mm culet and 3.1~mm base. However, if higher pressures are desired and high-precision sample preparation techniques~\cite{helm2020non} are available, smaller culets can be adapted to the pressure cell very easily by attaching diamonds with different shapes.

\subsection{End plates and screws}
Although G11, with a compressive strength of 434 MPa \cite{ASTMD695}, is one of the strongest plastic composite materials, the material is not suitable for holding fine screw threads due to the embedded glass fiber. Thus we use metal screws and end plates with threads to provide the compressive force. The end plates are at the ends of the cell, far away from the sample  (so that any heat induced there does not travel to the sample during  the $\sim 100$~ms magnet pulse), and the screws are placed on the edges of the cell body.  The screws and fixing plates are made of high strength and high-electrical-resistance grade 5 titanium alloy. The thickness of the G11 seats is calculated to compensate for the differential thermal contraction between Ti and G11, thus minimizing pressure drift between room and cryogenic temperatures.

\begin{figure}
\includegraphics[width=\linewidth]{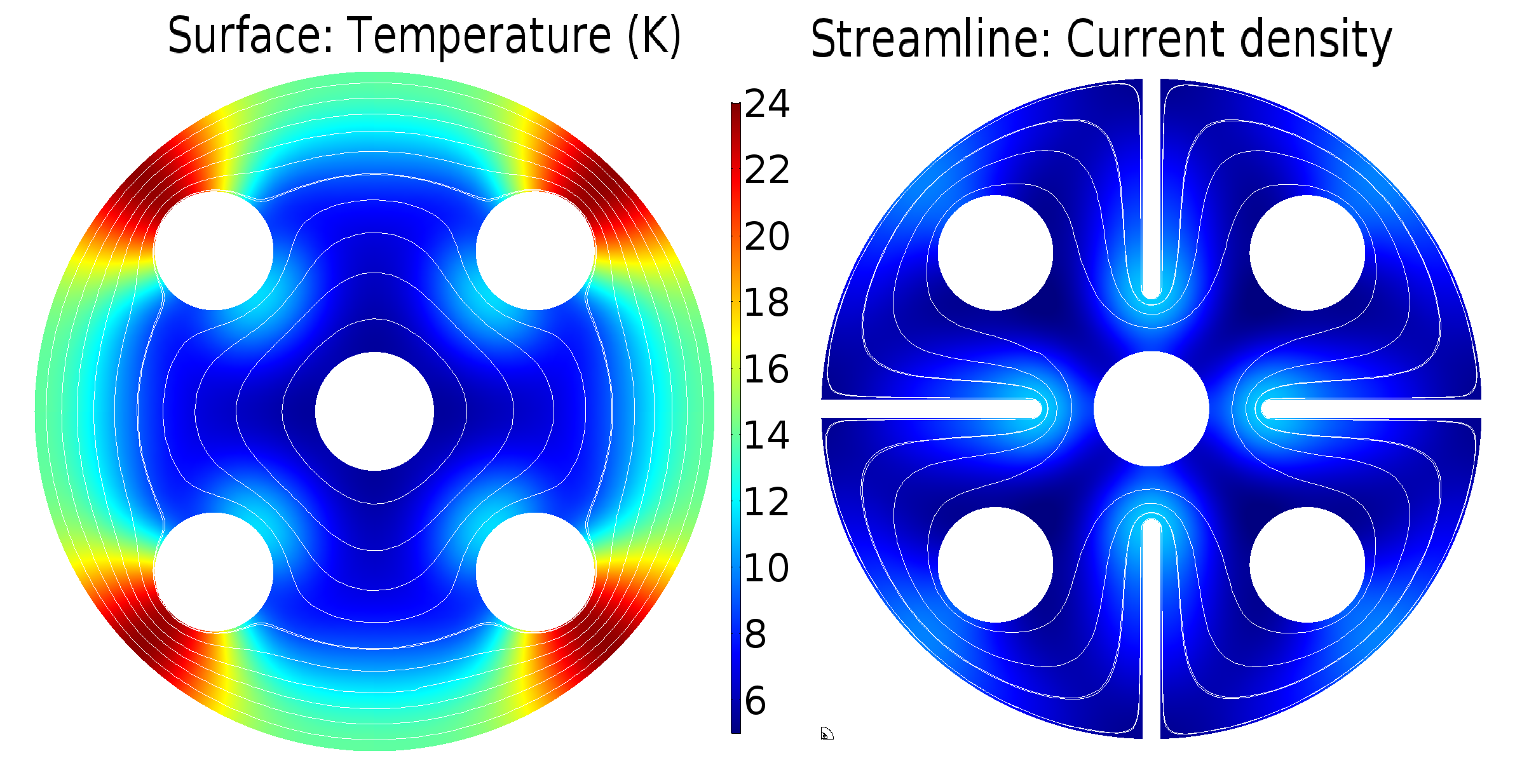}
\caption{\label{fig:endplate} Modeling of the Joule heating of the end plates using COMSOL Multiphysics. The temperature after the pulse is represented by the color scale and lines show the current flow. The initial temperature before the pulse is 4.2 K.}
\end{figure}

The Ti end plates, which have a large surface area perpendicular to the field, have slots to reduce eddy currents. We performed stress and Joule-heating evaluation of various designs using finite-element analysis, combined with information on the mechanical and electrical properties of several materials to refine the design. The heating by eddy currents was modeled using the COMSOL Multiphysics package. The modeling shows that the heating in the end plates from initial temperature of 4.2~K can be significantly reduced with strategically placed slots (Fig.~\ref{fig:endplate}). With the slots, the simulation gives an end-plate maximum temperature of 11.1~K after the pulse, with an accumulated Joule heating of 2.2~mJ; with no slots, the maximum temperature at the pulse end is 23.8~K, with 7.6~mJ accumulated energy. With coolant in the environment, both temperature rises can be much lower than the modeled maximum temperature. 

\subsection{Gaskets}
 
Conventional metal gaskets should be avoided in pulsed field at cryogenic temperatures, as they are ring-shaped and present a large area perpendicular to the field. They therefore provide an ideal path for eddy currents that is in close proximity to the sample, leading to considerable heating. By geometry, even a gasket with only a metal outer rim will have a large cross-section and will heat up the sample. Instead, we use a non-metallic compound gasket~\cite{lin2003amorphous}. The compound gaskets consist of an inner disk made of Boron powder mixed with epoxy and an outer ring made of Kapton. The diameter of the Boron part is 1~mm, a little bigger than the culet of the diamond, so that the diamonds press on the boron but the material between the diamonds still has space to flow. Kapton is used to provide radial support for the powdered region of the gasket. To make the gasket, boron is first mixed with epoxy (Epoxy Technology; part No.~353ND) in the mass ratio 3:2. A hole of 1~mm is drilled through the center of a Kapton rod, forming a tube. The mixture is then placed in the hole inside the Kapton tube and pressed using 100~lb of force. The mixture is then cured at $100^\circ$C for 24 hours within the tube. The cured mixture (and the tube) is cut into thick disks and the disks are polished to have the right thickness. With the current diameter of the powder-containing region and the culet size, we use a gasket thinner than $150~\mu$m. Then a $300~\mu$m hole is drilled through the center of the powder-containing region. With the central part press-cured, the gasket is used without pre-indentation, to prevent cracking. 
\subsection{Pressurization procedure}

The samples and the coil are prepared beforehand. The gasket with the sample hole is glued to the lower seat diamond. Then the samples and the coil are carefully placed into the sample hole and the sample space is filled with pressure medium. Next, the cell is temporally locked with the fixing screws. After that, the cell is  placed into a loading jig which utilizes thicker screws to apply the force. The pressure is monitored during gradual increase of the jig's load. After reaching the desired pressure, the cell is locked by the four fixing screws in the cell. 

\section{Performance tests}
\begin{figure}
\includegraphics[width=\linewidth]{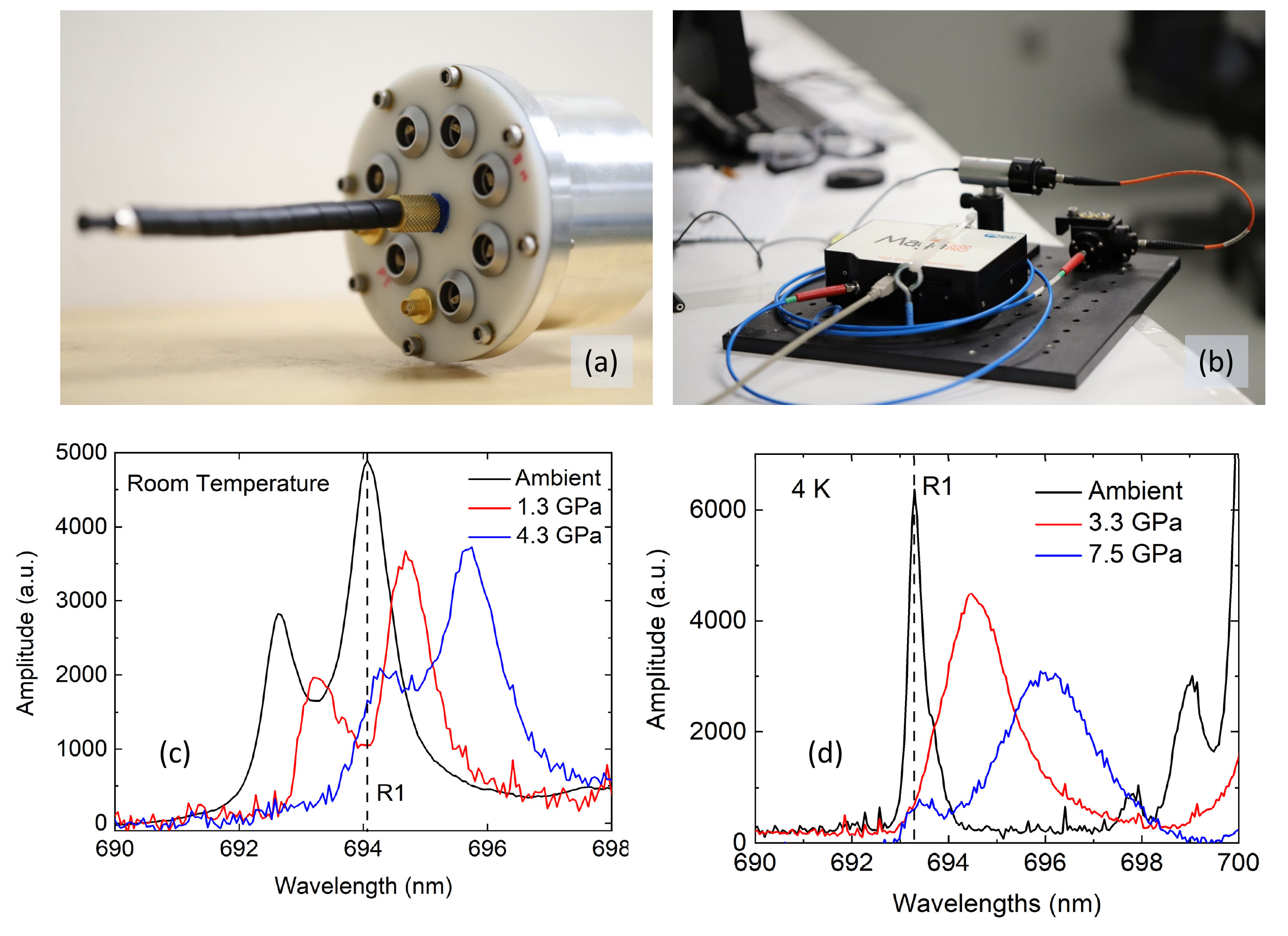}
\caption{\label{fig:pressure_deter} (a)~The 3D-printed probe top with optical fiber for ruby fluorescence. (b)~The spectrometer station. (c)~Example spectra from the cell at room temperature. (d)~Example spectra from the cell at low temperature.}
\end{figure}

%tests
We assembled the pressure cell and designed a custom probe to conduct high-pressure measurements in pulsed-field magnets. An optical fiber is fed to the pressure cell via the probe, so that the pressure can be monitored \textit{in situ} at low temperatures just before the magnet pulse using Ruby florescence. The hermetic connector bulkhead at the probe top is 3D printed via stereolithography~\cite{willis2020cryogenic}. The 3D-printed bulkhead forms a good vacuum seal and and can easily accommodate an arrangement of multiple electrical, fiber-optic, and radio frequency feedthroughs.  (Fig.~\ref{fig:pressure_deter}(a)).  We designed a portable pressure measuring station using an OZ Optics beam splitter and a compact spectrometer (Ocean Optics Maya 2000 Pro) on a $12" \times 12"$ optical breadboard  (Fig.~\ref{fig:pressure_deter}(b)). The pressure can be measured in real time by fitting the ruby fluorescence peak using a LabVIEW program. Reference ruby spectra at ambient pressure down to low temperature are obtained using bare ruby pieces attached to the optical fiber. The low temperature pressure is then calculated\cite{yamaoka2012ruby} using $P({\rm GPa})= 2.74\Delta\lambda$ with $\Delta \lambda$ being the wavelength difference in nm between the R1 peaks in the ambient and high-pressure spectra. Fig.~\ref{fig:pressure_deter}(d) shows the spectrum at 7.5~GPa and 4.0~K, with a sample space diameter of $300~\mu m$. The factor limiting the maximum pressure comes from the fixing screws; see Supplementary Material  Section~1 for details. The fact that the anvil seats and the gasket can survive to higher pressures than the screws means that the maximum pressure can be increased by using smaller culets or by adding more screws to the design.

\begin{figure}
\includegraphics[width=\linewidth]{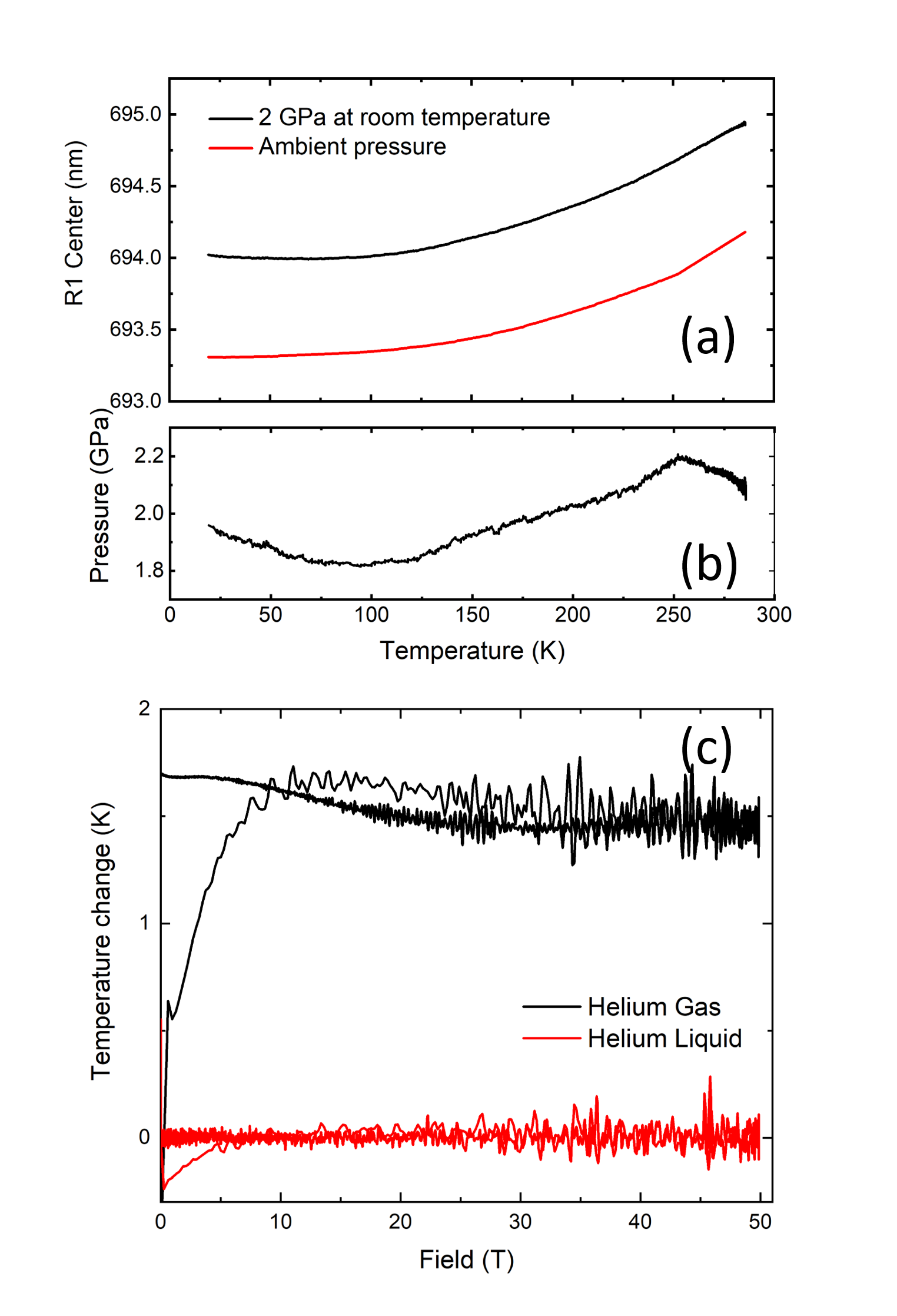}
\vspace{-8mm}
\caption{\label{fig:cell_test} (a) and (b) The change in pressure when the cell is cooled from room temperature to low temperature. The position changes of ruby fluorescence peak R1 are shown in (a)~for ambient pressure and for a pressure of 2~GPa on the cell.  (c)~The change in sample temperature in a 50~T pulse starting at 4.0~K.}
\end{figure}

In designing the cell, we calculated the length of the G11 body, the thickness of the end plates and the length of the Ti screws so that the difference in thermal contraction between the Ti and G11 components is compensated. We then tested the pressure change between room temperature and low temperature. With a pressure $\approx 2$~GPa, the pressure variation between room- and low temperature is $< 10\%$ [see Fig.~\ref{fig:cell_test}(a)]. 

One of the main concerns in a pulsed-field measurement is sample heating. The cell is designed to have minimal heating during the pulse, especially in the areas close to the sample. To test for sample heating during the pulse, a miniature bare-chip Cernox thermometer was placed on the side of the diamond. The thermal conductivity of diamond is five times bigger than silver at room temperature and becomes even better at low temperatures~\cite{wei1993thermal}. Thus the temperature of the Cernox can be treated as a good representation of the sample temperature. The resistance of the Cernox is then measured using a high frequency AC lock-in technique during the field pulse. Fig.~\ref{fig:cell_test}(b) shows that the eddy-current heating of the sample in a 50~T magnet pulse is about 1.5~K in Helium gas and unnoticeable in Helium liquid at 4~K. 

\begin{figure}
\includegraphics[width=\linewidth]{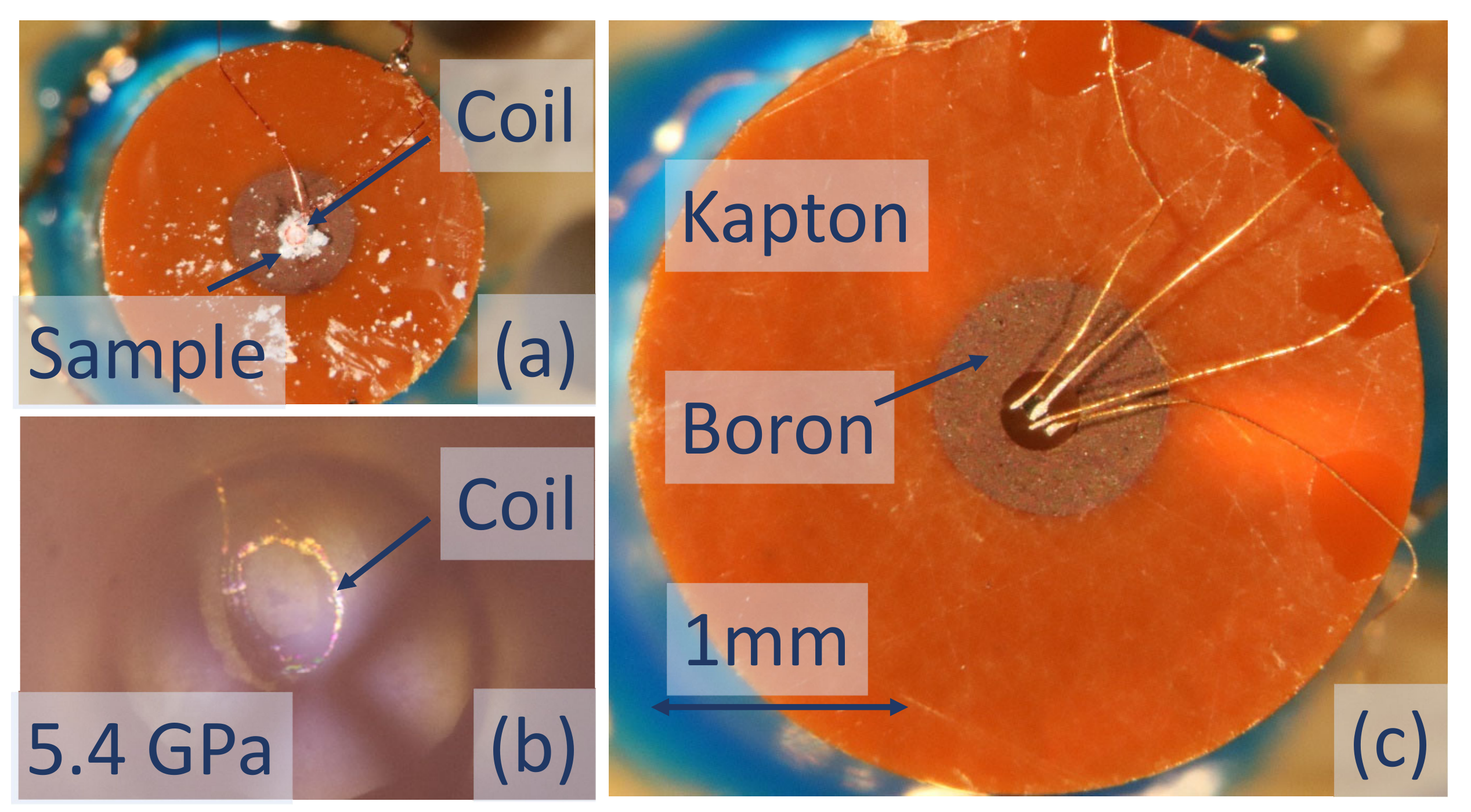}
\caption{\label{fig:PDO} (a) The PDO setup on a gasket, with the sample inside the sample space. (b) The PDO coil still retains its shape at a pressure of 5.4~GPa. The photo is taken through the diamond. (c)The transport setup on a gasket.}
\end{figure}

%PDO
The cell has a relatively large sample space for a diamond anvil cell; the diameter is $300~\mu$m and the depth is $120~\mu$m. The large sample space permits the use of a variety of measurement techniques. The proximity-detector-oscillator (PDO) technique has been tested successfully with this cell~\cite{Sun2020PDO}. The PDO was designed to be used in pulsed magnetic fields to measure skin depth, penetration depth and magnetic susceptibility of conductors, superconductors and insulators, respectively. The measurement is contactless and can work with a coil of just 4 turns wound around the sample. This technique is especially suitable for the limited space of a pressure cell and for samples that are difficult to connect electrical leads to or which are insulators. Miniature coils with an outer diameter of $140~\mu$m are made with $9~\mu$m diameter copper wires (Fig.~\ref{fig:PDO}(a)). Fig.~\ref{fig:PDO} (b) shows that the wire survives and the original shape of the coils remains largely unchanged under a pressure of 5.4~GPa. With careful arrangement of the sample and the electrical leads, transport measurements can also be carried out. The sample is connected to electrical leads using silver paste and is surrounded by the pressure medium. In those experiments, we use Daphne 7373 as pressure medium.  One of the configurations is shown in Fig.~\ref{fig:PDO}(c).

\section{Conclusions}

A composite pressure cell and gaskets have been built to avoid the eddy-current Joule heating induced by pulsed magnetic fields. A pressure of 8~GPa is achieved in a $300~\mu$m diameter sample space. The pressure cell is able to host bulk or powdered samples and is applicable to transport and PDO measurements. Sample heating is indiscernible at cryogenic temperatures. The change in pressure between room- and low temperatures is small. Moreover, the design enables quasi-mass production of cell components, so that a fast turn-around rate is achievable, desirable for pulsed-field facility users with limited time on-site. This is aided by the relatively easy loading of the cell, which does not require a hydraulic press.

\section*{Supplementary Material}
See the supplementary material for details on maximum pressure limits, the pressurization procedure, and additional pulsed field heating tests. 

\begin{acknowledgments}
This work was supported by the US DoE Basic Energy Science Field Work Project {\it Science of 100 T}, and carried out at the National High Magnetic Field Laboratory, which is funded by NSF Cooperative Agreements DMR-1644779, the State of Florida and U.S. DoE.

\end{acknowledgments}

\section*{Data Availability Statement}
The data that support the findings of this study are available from the corresponding author upon reasonable request.

\appendix
\section{Limiting factors for the maximum pressure}

The failure modes for the screws in the DAC are calculated using Federal Standard Fastener Calculations \cite{NASA5020} and shown in Supplementary Table 1. These are general loading limits derived from von Mises failure criteria; since the torque is never applied to the screws while they are under full load, many failure modes can be simplified to pure shear/uniaxial stress. Although the screws are long and unguided in the section where the anvils are located, the maximum bending moment is not calculated since there is limited shear across the DAC. All calculations utilize G5 Titanium shear and tensile yield strengths at room temperature, in which thermal loading or fatigue failure are not considered. In fact, Titanium is typically stronger at low temperatures.

\begin{figure}[h]
\includegraphics[width=\linewidth]{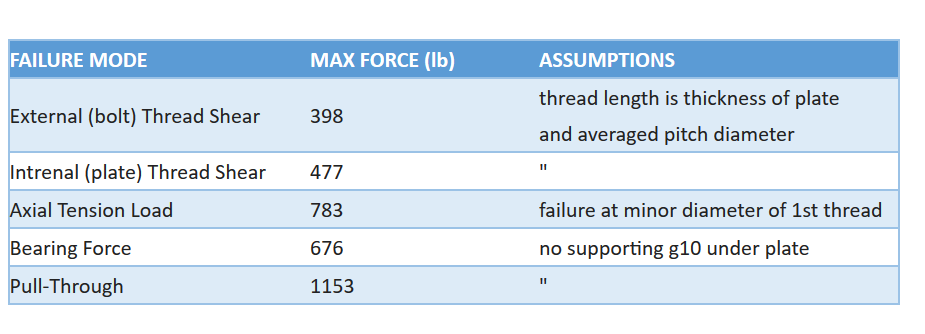}
\label{fig:1} \textbf{Supplementary Table 1}. Maximum force for each failure mode. The forces are calculated for one screw. 
\end{figure} 

 Consequently at failure load, the uniform pressure with a pair of 0.8 mm-culet diamonds is $\approx 14$~GPa. The general rule of thumb for fasteners to avoid thread shear failure is to have a thread length $1.5-2$ times the diameter of the bolt. The thread length in the Ti end plate is 0.8 times the bolt diameter so the threads are the most vulnerable failure points.
 
 Since G11 laminate sheet has a compressive strength of 434~MPa \cite{ASTMD695}, than for anvils with 3.1 mm base and 0.8 mm culet the G11 seats will have a pressure limit of 6.5 GPa assuming uniform pressure across the culets. 
 In real experiment the pressure peaks at the center and drops off toward the edge of the culet. We reach 12 GPa the center of the cell with the seats in a loading jig. We examine the seats after the experiments, and we find no indentation on the seats. This shows that the seats can work up to at least 12 GPa with 0.8 mm culets.

The gasket fails in a manner that is quite complex. As it gets compressed between two anvils, it flows out until the friction between the gasket and anvils stops the flow. In case of a soft epoxy plus hard powder composite, the epoxy may get squeezed out and the powder may get jammed. The Kapton ring also helps, but it clearly gets stretched from a flat into a conical form. At high pressure, the anvils deform as well into concave shape as the pressure in the center is higher than at the perimeter.  A work by others\cite{merkel2005x} shows a very similar $150~\mu$m thickness Kapton with boron-epoxy gasket being used up to around 65~GPa. Thus the gasket will outlast the titanium threading.

\section{Time evolution of an exemplary pressure loading}
\begin{figure}[h]
\includegraphics[width=\linewidth]{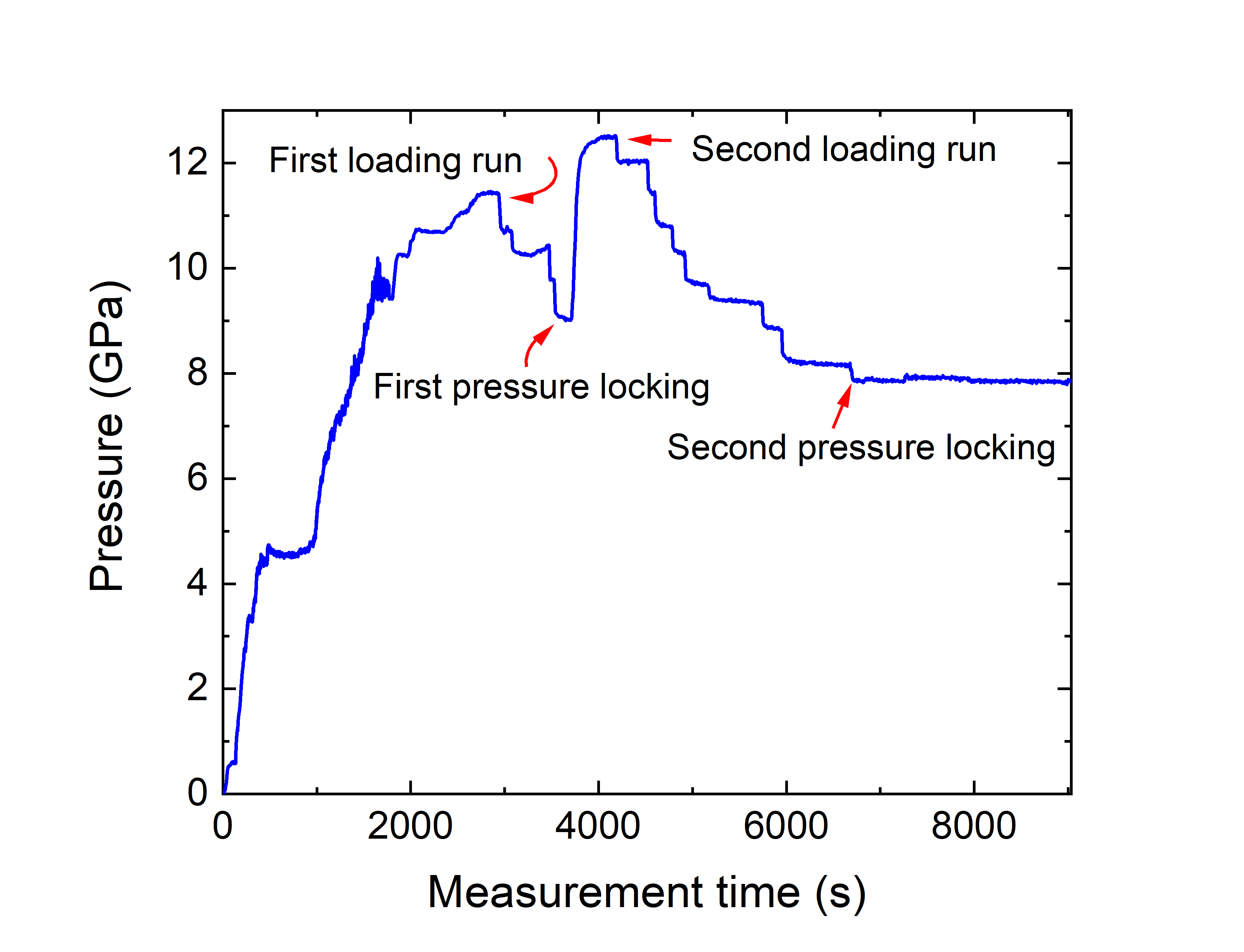}
 \raggedright{\textbf{Supplementary Figure 1}. Pressure versus measurement time. The final pressure is approached in several steps in a slow fashion. In the first loading run, the pressure is applied with a loading jig to 11.4 GPa and the fixing screws are tightened. The jig load is then partially released and the pressure dropped to 9 GPa. The second loading run and pressure locking repeat similar steps and finally 8~GPa can be stabilized after completely releasing the jig load.   The pressure value is calculated by fitting the R1 peak in the spectrum.}
\end{figure}

In a routine pressure loading, the cell is first locked by the fixing screws and then placed in a  loading jig. The pressure is loaded gradually and the pressurization is stopped when the pressure is a few GPa higher than the desired final value. Then the pressure is locked by the fixing screws. 

Experiments like the one shown in Supplementary Fig. 1 demonstrate that the G11 seats and the gaskets can work up to at least 12 GPa. After the fixing screws are locked, the cell is removed from the loading jig, and the pressure returns to about 8~GPa due to the stretching of the fixing screws. This pressure can be maintained for several days until the measurement in pulsed fields is finished.

\section{Further data on heating}
\begin{figure}[h]
\includegraphics[width=\linewidth]{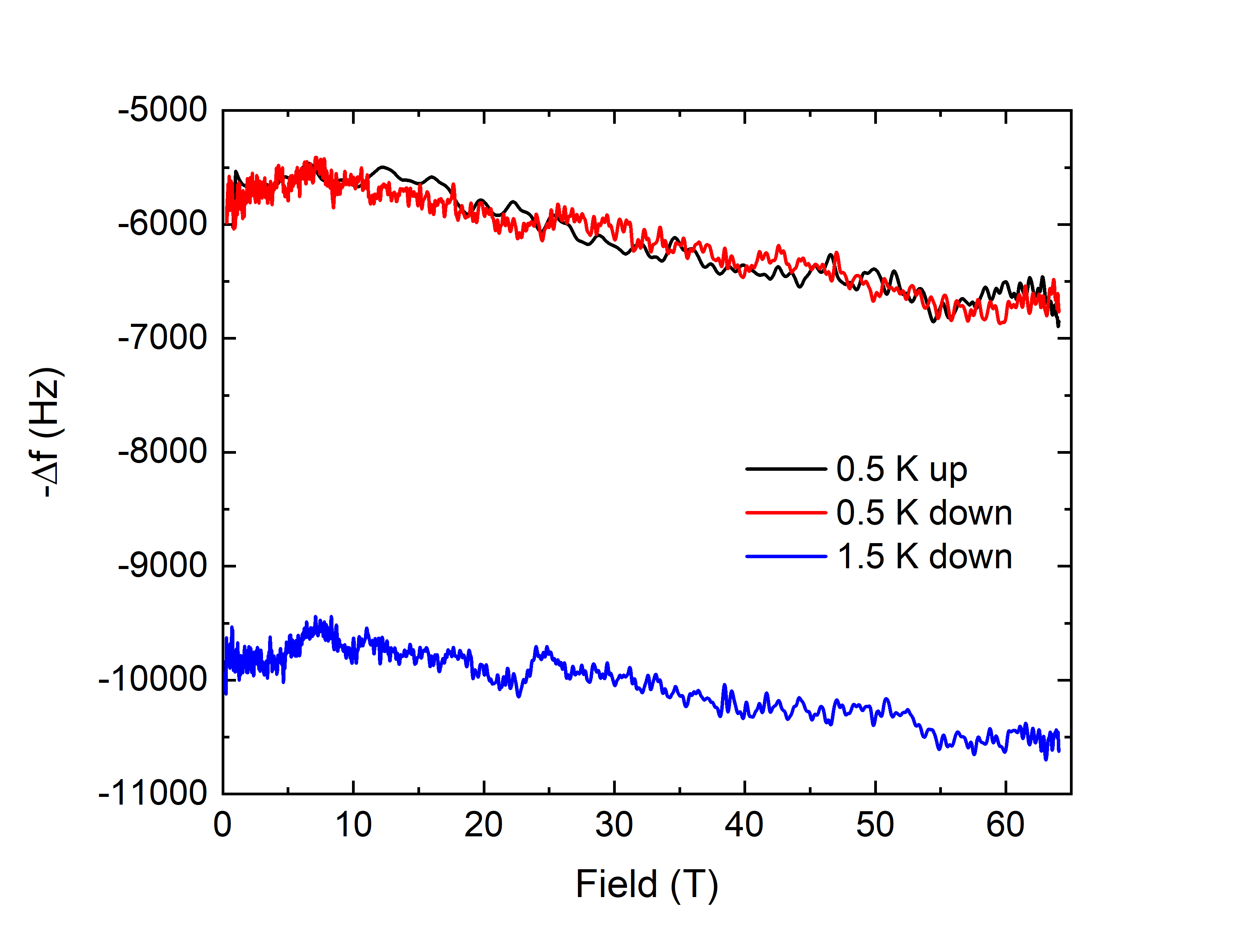}
 \raggedright{\textbf{Supplementary Figure 2}.  The data are taken from a PDO measurement at 5.4~GPa. The frequency change is proportional to the change in magnetic susceptibility of the sample.}
\end{figure}

Supplementary Fig. 2 shows that the 0.5~K data from the up-sweep and the down-sweep parts of the magnetic field pulse agree really well: no significant hysteresis is noticeable. The 1.5~K data are shifted in frequency because the sample susceptibility has a large temperature dependence in this temperature range. This significant temperature dependence and the absence of hysteresis in the 0.5 K data combine to show that heating during the pulse is very small.

%\nocite{*}
%\bibliography{Pcell}% Produces the bibliography via BibTeX.

\end{document}